%% file: CFT-09-008_temp.tex
\pdfoutput=1
\documentclass[11pt,twoside,a4paper,pdftex,cmspaper,final,collab]{cms-tdr}

\begin{document}\cmsNoteHeader{CFT-09-008}
%
%
%

%
%
\hyphenation{env-iron-men-tal}
\hyphenation{had-ron-i-za-tion}
\hyphenation{cal-or-i-me-ter}
\hyphenation{de-vices}
%
%
\RCS$Revision: 1.52 $
\RCS$Date: 2010/01/13 01:47:50 $
\RCS$Name:  $
\input{ptdr-definitions}
\cmsNoteHeader{09-008}
\title{Commissioning of the CMS Experiment and the Cosmic Run at Four Tesla}

\address[cern]{CERN}
\author[cern]{CMS Collaboration}

\date{\today}

\abstract{
The CMS Collaboration conducted a
month-long data-taking exercise known as the Cosmic Run At Four Tesla
in late 2008 in order to complete the commissioning of the experiment for
extended operation. The operational lessons resulting from this exercise
were addressed in
the subsequent shutdown to better prepare CMS for LHC beams in 2009.
The cosmic data collected have been invaluable to study
the performance
of the detectors, to commission the alignment and calibration techniques,
and to make several cosmic ray measurements.
The experimental setup, conditions, and principal achievements from
this data-taking exercise are described along with a review of the
preceding integration activities.
}

\hypersetup{%
pdfauthor={D. Acosta},%
pdftitle={Commissioning of the CMS Experiment and the Cosmic Run at Four Tesla},%
pdfsubject={CMS},%
pdfkeywords={CMS, CRAFT, detector, computing, DAQ, trigger, calorimeter,
muon}
}

\maketitle 

\section{Introduction}

The primary goal of the Compact Muon Solenoid (CMS) experiment \cite{JINST} is
to explore particle
physics at the TeV energy scale, exploiting the proton-proton
collisions delivered by the Large
Hadron Collider (LHC) at CERN \cite{lhc}.
The complexity of CMS, like that of the other LHC experiments,
is unprecedented. Therefore, a focused and comprehensive
programme over several years, beginning with the commissioning of
individual detector subsystems and transitioning to the commissioning
of experiment-wide operations, was pursued to bring CMS into full
readiness for the first LHC beams in September 2008.
After the short period of beam operation
the CMS Collaboration conducted a
month-long data-taking exercise known as the Cosmic Run At Four Tesla (CRAFT)
in late 2008. In addition to commissioning the experiment
operationally for an extended period,
the cosmic muon dataset collected during CRAFT has proven invaluable
for understanding the performance of the CMS experiment as a whole.

The objectives of the CRAFT exercise were the following:
\begin{itemize}
\item Test the solenoid magnet at its operating field
($3.8\,$T), with the
CMS experiment in its final configuration underground;
\item Gain experience operating CMS continuously for one month;
\item Collect approximately 300 million cosmic triggers for
performance studies of the CMS subdetectors.
\end{itemize}
The CRAFT exercise took place from October 13 until November
11, 2008, and these goals were successfully met.

This paper is organized as follows.
Section~\ref{sec:description} describes the detectors comprising CMS while
Section~\ref{sec:install} describes the installation and
global commissioning programme prior to CRAFT. The experimental setup for CRAFT
and the operations conducted are described
in Sections~\ref{sec:setup} and \ref{sec:operations}, respectively,
and some of the analyses made possible by the CRAFT dataset are
described in Section~\ref{sec:performance}.

\section{Detector Description}
\label{sec:description}

\begin{figure}
\begin{center}
\includegraphics[width = 350pt]{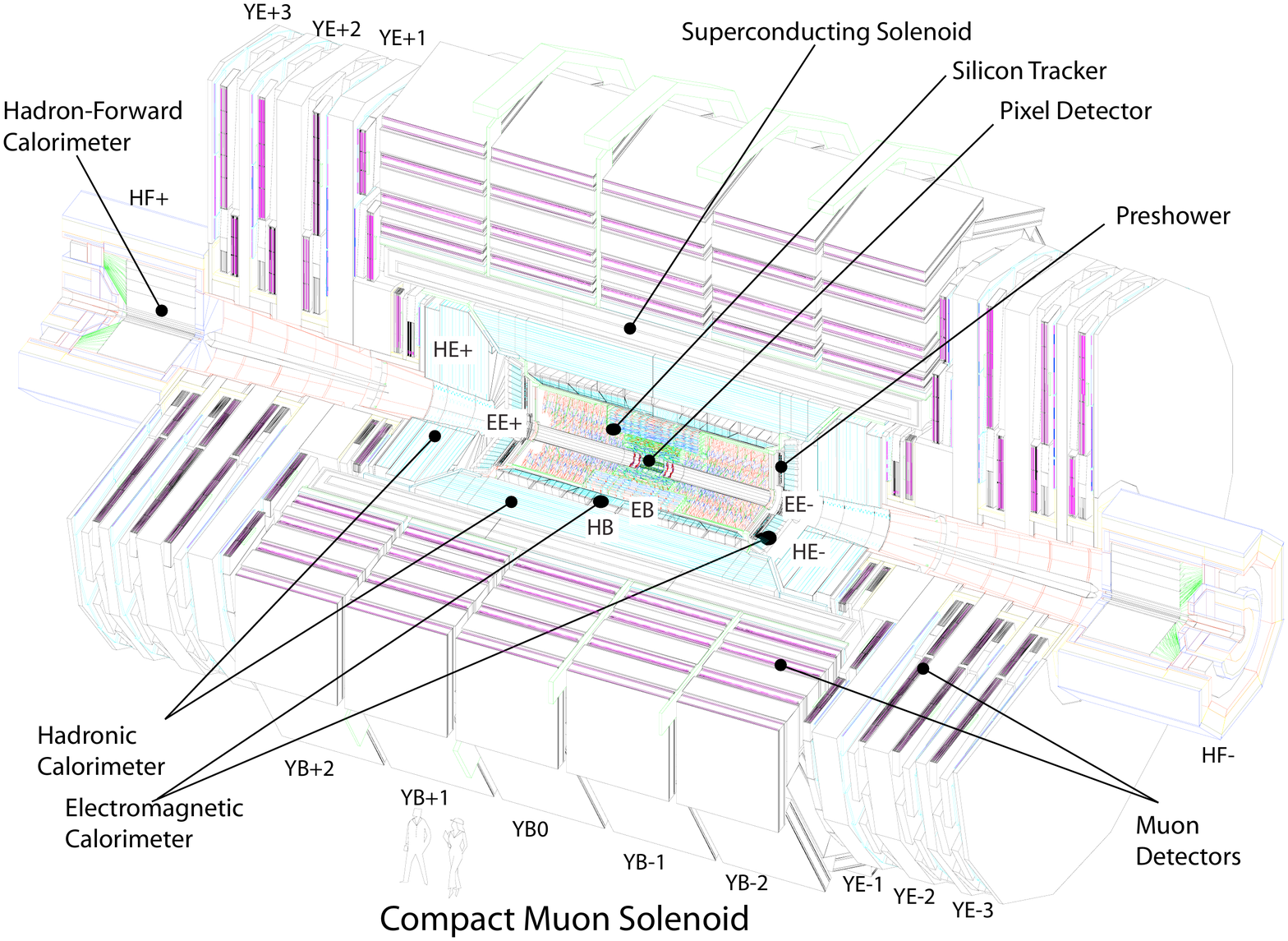}     
\caption{
General view of the CMS detector. The major detector components are indicated, together with the acronyms for the various CMS construction modules.
 \label{fig:CMSfigure} }
\end{center}
\end{figure}

A detailed description of the CMS experiment,
illustrated in Fig.~\ref{fig:CMSfigure},
can be found elsewhere~\cite{JINST}.
The central feature of the CMS apparatus is a superconducting
solenoid of $6\,$m internal diameter, $13\,$m length, and designed to
operate at up to a field of $4\,$T.
The magnetic flux generated by the solenoid is returned via the
surrounding steel return yoke---approximately $1.5\,$m thick, $22\,$m long, and
$14\,$m in diameter---arranged as a 12-sided cylinder closed at each end by
endcaps. To facilitate pre-assembly of the yoke and the installation
and subsequent maintenance of the detector systems, the barrel yoke is
subdivided into five wheels (YB$0$, YB$\pm 1$, and YB$\pm 2$, as
labeled in Fig.~\ref{fig:CMSfigure}) and each
endcap yoke is subdivided into three disks (YE$\pm 1$, YE$\pm 2$, and YE$\pm 3$).
Within the field volume are the
silicon pixel and strip trackers, the lead tungstate crystal electromagnetic
calorimeter (ECAL), and the brass-scintillator hadronic calorimeter
(HCAL).
Muons emerging from the calorimeter system are measured in
gas-ionization detectors embedded in the
return yoke.

CMS uses a right-handed coordinate system, with the origin at the
nominal interaction point, the $x$-axis pointing to the centre of the
LHC, the $y$-axis pointing up (perpendicular to the LHC plane), and
the $z$-axis along the anticlockwise-beam direction.
The polar angle, $\theta$, is measured from the positive $z$-axis,
and the pseudorapidity $\eta$ is defined as $\eta=-\ln{\tan{(\theta/2)}}$.
The azimuthal angle, $\phi$, is measured in the $x$-$y$ plane.

Charged particles are tracked within the pseudorapidity range
$|\eta|<2.5$.
The silicon pixel tracker consists of
$1440$ sensor modules containing a total of 66~million
$100\times150$~$\mu$m$^2$ pixels. It is
arranged into three
$53.3\,$cm long barrel layers and two endcap disks at each end. The
innermost barrel layer has a radius of $4.4\,$cm, while the other two
layers are located at radii of $7.3\,$cm and $10.2\,$cm. The endcap disks
extend in radius from about $6\,$cm to $15\,$cm and are located at
$\pm 34.5\,$cm and $\pm 46.5\,$cm from the interaction point along the beam axis.
The silicon strip tracker consists of $15\,148$
sensor modules containing a total of
$9.3$ million strips with a
pitch between 80 and 180$\,\mu$m.
It is $5.5\,$m long and $2.4\,$m in diameter,
with a total silicon surface area of $198\,$m$^2$.
It is constructed from six subassemblies: a four-layer inner barrel (TIB), two
sets of inner disks (TID) comprising three disks each, a six-layer
outer barrel (TOB), and two endcaps
(TEC) of nine disks each.

The ECAL is a fine grained hermetic calorimeter consisting of
$75\,848$ lead tungstate (PbWO$_4$) crystals that provide
fast response, radiation tolerance, and excellent energy
resolution. The detector consists of a barrel region, constructed from
36 individual supermodules (18 in azimuth per half-barrel), extending to
$|\eta|=1.48$, and two
endcaps, which provide coverage up to $|\eta| = 3.0$.
The crystals in the barrel have a transverse cross-sectional area at the
rear of $2.6\times
2.6\,$cm$^2$, corresponding to
$\Delta\eta\times\Delta\phi = 0.0174\times 0.0174$, and a longitudinal
length
of 25.8 radiation lengths. The crystals in the endcap
have a transverse area of $3\times 3\,$cm$^2$ at the rear and a
longitudinal length of
24.7 radiation lengths.
Scintillation light from the crystals is detected by avalanche
photodetectors in the barrel region and by vacuum phototriodes
(VPT) in the endcaps. A preshower detector
comprising two consecutive sets of lead radiator followed by silicon strip
sensors was mounted in front of the endcaps in 2009, after the CRAFT
period, and has a
thickness of three radiation lengths.

The HCAL barrel (HB) and endcaps (HE) are sampling
calorimeters composed of brass and scintillator plates with coverage
$|\eta|<3.0$. Their thickness varies from 7 to 11
interaction lengths depending on $\eta$;
a scintillator ``tail catcher'' placed
outside of the coil at
the innermost muon detector extends the instrumented thickness to more
than 10
interaction lengths everywhere. In the HB,
the tower size is $\Delta\eta\times\Delta\phi = 0.087\times 0.087$.
Each HB and HE tower has 17 scintillator layers
except near the interface of HB and HE.
The scintillation light is converted by
wavelength-shifting fibres embedded into the scintillator tiles, and
is then channeled to hybrid photodiodes (HPD) via clear optical fibres.
Each HPD collects signals from up to 18 different HCAL towers.
The Hadron Outer (HO) calorimeter comprises
layers of scintillators
placed outside the solenoid cryostat to catch the energy leaking out of
the HB. Its readout
is identical to that of the HB and HE.
Quartz fibre and iron forward
calorimeters (HF), read out by photomultipliers,
cover the $|\eta|$ range between 3.0 and 5.0, which corresponds to
the conical central bore of each endcap yoke.

Three technologies are used for the detection of muons:
drift-tubes (DT) in the central
region ($|\eta|<1.2$), cathode strip chambers (CSC) in the endcaps
($0.9<|\eta|<2.4$), and resistive
plate chambers (RPC) throughout barrel and endcap ($|\eta|<1.6$). The DT system
comprises 250  chambers mounted onto the five wheels of the
barrel yoke and
arranged into four concentric ``stations'' interleaved with the steel yoke plates.
Each chamber is built from a
sandwich of 12 layers of drift tubes with $4.2\,$cm pitch,
and is
read out with multiple hit capability.
Eight layers have wires along $z$ and measure the $\phi$ coordinate;
four layers have wires perpendicular to the $z$-axis and measure
$z$ (except for the outermost DT station where there are no $z$ measuring layers).
The CSC system is made of 468 chambers mounted on the faces of the
endcap disks, so as to give four stations perpendicular to the beam pipe
in each endcap.
Each chamber has six
cathode planes segmented into narrow trapezoidal strips projecting radially
from the beam line, and anode wires aligned perpendicularly to the
strips (wires for the highest $|\eta|$ chambers on YE$\pm
1$ are tilted by $25^\circ$ to compensate for the Lorentz angle).
The barrel RPC system is mounted in the same pockets in the yoke
wheels as the DT system, but with six concentric layers of
chambers. Each endcap RPC system consists of three layers mounted on
the faces of the yoke disks. Each RPC chamber contains two gas gaps of
$2\,$mm thickness, between which are sandwiched readout strips that
measure the $\phi$ coordinate. The gaps work in saturated avalanche mode.
The relative positions of the different elements of the muon system
and their relation to reference elements mounted on the silicon strip tracker
are monitored by a sophisticated alignment system.

A system of beam radiation monitors installed along the beam line gives
online feedback
about the beam structure and about radiation conditions within the
experimental cavern \cite{BRM,BCM}. The main components are
radio frequency (RF) pickups located $\pm 175\,$m from the
interaction point, segmented scintillator rings mounted  on both
faces of the HF calorimeters, and diamond sensors installed very close
to the beam pipe at distances of $\pm 1.8\,$m and $\pm 14.4\,$m.
Signals from
the diamond beam condition monitors are used to protect the tracking
detectors from potentially dangerous beam backgrounds. In severe
pathological conditions, they are capable of triggering an abort of
the LHC beams.

Only two trigger levels are employed in CMS. The Level-1 trigger is
implemented using custom hardware processors and is designed to reduce the
event rate to at most $100\,$kHz during LHC operation using coarse information from the
calorimeters, muon detectors, and beam monitoring system. It operates
with negligible
deadtime and synchronously with
the LHC bunch crossing frequency of $40\,$MHz.
The High Level Trigger
(HLT) is implemented across a large cluster
of the order of a thousand
commercial computers,
referred to as the event filter farm \cite{EVFcommission}, and provides further
rate reduction to $\mathcal{O}(100)\,$Hz using filtering software applied to
the full granularity data acquired from all detectors.
Complete events for the HLT are assembled from the fragments sent from each detector
front-end module through a complex of switched networks and ``builder
units'' also residing in the event filter farm.
The event filter farm is
configured into several ``slices'', where each slice has an
independent data logging element (``storage manager'') for the storage of
accepted events.

\section{CMS Installation and Commissioning Programme prior to CRAFT}
\label{sec:install}

The strategy for building the CMS detector is unique among the four major
experiments for the LHC at CERN. The collaboration decided from the beginning
that assembling the large units of the detector would take place in a surface hall
before lowering complete sections into the underground experimental
cavern.
This philosophy allowed the CMS construction effort to be completed on
time despite delivery of the underground cavern late in the schedule,
as a result of
civil-engineering works that were complicated by the geology of
the terrain. Another goal was to
minimize underground assembly operations which would inevitably have
taken more time and would have been more complex and
risky in the confined space of the cavern. Future access to
the  inner parts of the detector is also made easier.
As construction and assembly
progressed above ground, it became clear that there would be a valuable
opportunity for system integration and commissioning on the
surface.

\subsection{The 2006 Magnet Test and Cosmic Challenge}
\label{sec:mtcc}

The large solenoid of CMS was first fully tested while it was in the
surface assembly
hall during August--November 2006. This provided  the opportunity to test
the integration of major
components of the experiment before lowering them into the underground
experimental cavern,
and slices of the major detector
subsystems were prepared to record data concurrently with this test.
The exercise, called the
Magnet Test and Cosmic Challenge (MTCC), provided important
commissioning and operational experience, and was the precursor of
the CRAFT exercise described in this paper. The magnetic field was increased
progressively
up to its maximum operating value of $4.0\,$T, and fast discharges were
commissioned such that 95\% of the operating current of $19\,140\,$A
(corresponding to $2.6\,$GJ of stored energy) could be dumped in a
time span of about 10 minutes. Distortions of the yoke during the
testing were monitored by the muon alignment
system \cite{Linksystem}, which was installed in one endcap and in
an azimuthal slice of the barrel across all wheels.
After the successful completion of testing in the surface hall in
2006, the magnet and its main ancillary systems were moved to
their final positions in the
service and experimental caverns and nearby surface
installations.

Concurrently with the first phase of the 2006 magnet test,
about 7\% of the muon detection
systems, 22\% of HCAL, 5\% of ECAL, a pilot silicon strip tracker
(about 1\% of the scale of the complete tracker), and the global trigger and data acquisition were successfully
operated together for
purposes of globally commissioning the experiment and for collecting
data to ascertain the detector performance. For the second phase of the
exercise, the
ECAL and pilot tracker were removed and the central magnetic field was
mapped with a precision of better than $0.1\%$, using a specially designed
mapping carriage employing Hall probes mounted on a rotating
arm \cite{MTCCmap}, for several operating fields of the magnet. These maps are now used
 in the offline simulation and reconstruction software.
In total, approximately 200 million cosmic
muon events were recorded for purposes of calibration, alignment, and
detector performance studies using this slice of the experiment while
on the surface.
The conclusion of MTCC coincided with the
start of the installation into the underground cavern.

The MTCC was an opportunity to uncover
issues associated with operating the
experiment
with the magnetic field at its design value.
One effect
seen was the susceptibility of the
hybrid photodetectors (HPD) used to read out the scintillation
light from the HCAL: the noise rate from these devices depends on the
magnetic field, and is maximal in the range 1--2$\,$T. At the design value
of $4\,$T the noise rate was found to be acceptable for the barrel and endcap
compartments of HCAL; but for the Hadron Outer ``tail catcher'' in the
barrel, whose HPDs are mounted in pockets in the return yoke where the
magnitude of the field is lower, the noise rate was
unacceptably high and the tubes had to be repositioned.

\subsection{Installation of CMS Components Underground}

The heavy elements of CMS began to be lowered into the experimental
cavern in November 2006, starting with the forward calorimeters and
continuing
shortly thereafter with the $+z$ endcap disks and barrel
wheels, complete with muon detectors and services.
The central yoke wheel (YB$0$), which houses the cryostat, was lowered
in February 2007, and by January 2008
the last heavy elements of the $-z$ endcap were successfully lowered
into the cavern.

The campaign to connect services for the detectors within the central portion of CMS
included the installation of more than $200\,$km of cables and optical
fibres (about 6000 cables). Additionally, more than
$20\,$km of cooling pipes (about 1000 pipes) were installed. The
whole enterprise took place over a 5 month period and required more
than $50\, 000$ man-hours of effort.
The cabling of the silicon strip tracker was completed in March 2008,
and its cooling was operational by June 2008.
In the same month, the central beam pipe, which is made of beryllium,
was installed and baked out (heated to above $200\,^\circ$C while under vacuum
for approximately a week).

The silicon pixel tracking system and the endcaps of the ECAL were
the last components to be installed, in August 2008.
The mechanics and the cabling of the pixel system have been designed to
allow relatively easy access or replacement if needed.
The preshower
detector for the endcap electromagnetic calorimeter was the only major
subsystem not installed prior to the 2008 LHC run and the CRAFT
exercise.  It was installed in March 2009.

\subsection{Global Run Commissioning Programme}

\begin{figure}
\begin{center}
\includegraphics[width = 450pt]{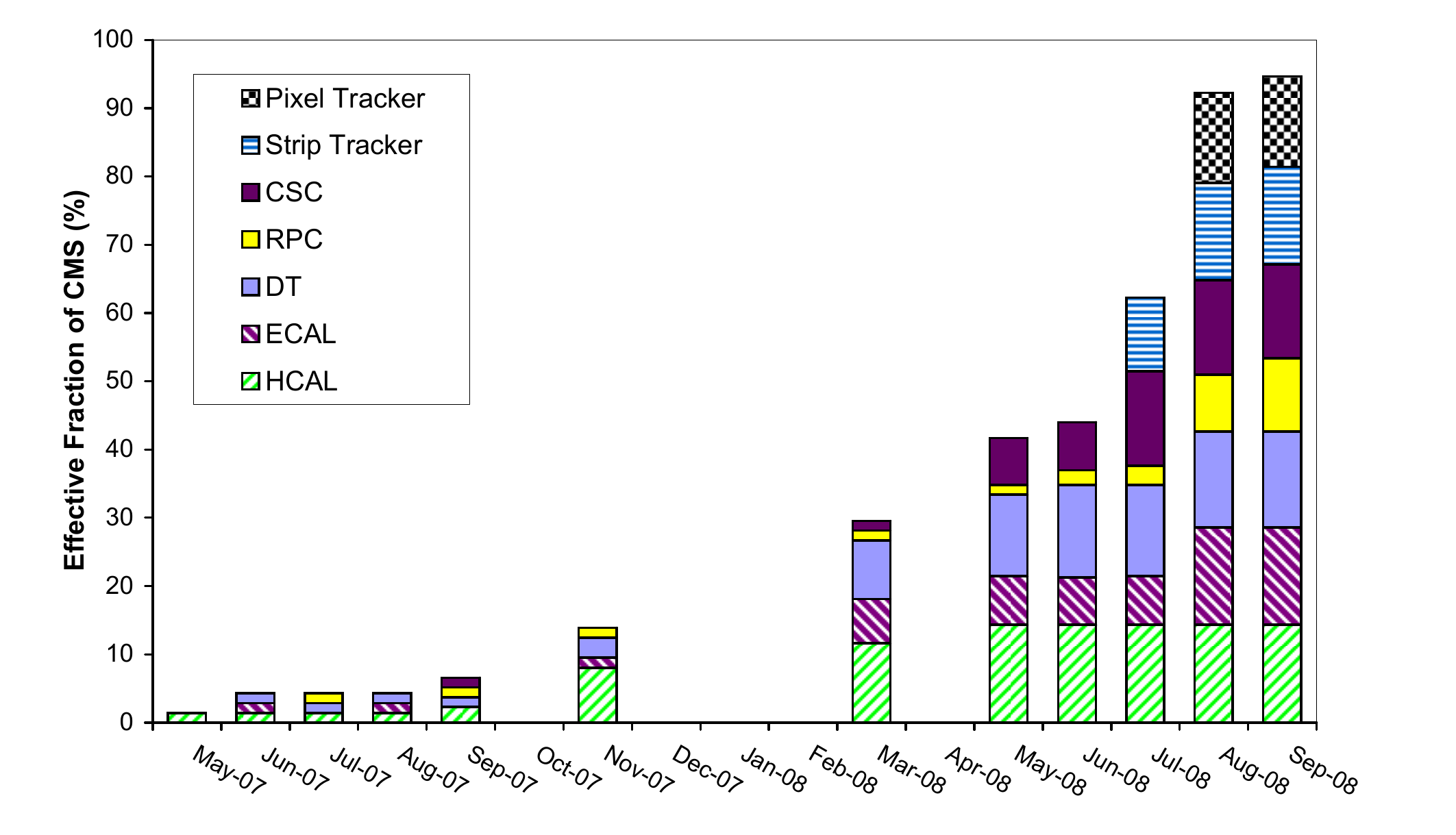}     
\caption{
Effective fraction of the CMS experiment participating in the 2007 and
2008 global run campaigns as a function of time. The fraction of each
of the seven major detector systems is represented by a bar with a
length of up to ${1\over 7}\cdot$100\%. Only one RPC endcap was missing by September 2008.
\label{fig:GRscale} }
\end{center}
\end{figure}

A series of global commissioning exercises
using the final detectors and electronics installed in the underground
caverns,
each lasting 3--10
days and occurring monthly or bimonthly, commenced in May 2007 and
lasted until the experiment was prepared for LHC beams, by the end of August 2008.
These ``global runs'' balanced the need to continue installation and
extensive detector subsystem commissioning
with the need for global system tests.
The scale of the global runs is illustrated in Fig.~\ref{fig:GRscale},
which shows, as a function of time, the effective fraction of each of
the seven major detector systems participating in the run
(excluding the ECAL preshower system).
Generally, the
availability of power, cooling, and gas limited the initial scope of the
commissioning exercises; by the time of the November 2007
global run, however, these services were widely available.

Many detector subsystems were available in their entirety for
global commissioning by  May 2008, and thus  a series of
four week-long exercises, each known as a Cosmic
RUn at ZEro Tesla (CRUZET), were conducted to accumulate sizable
samples of cosmic muon events from which to study the overall detector
performance.
Notable for the third CRUZET exercise, in July 2008,
was the introduction of the silicon strip tracker into the data-taking
(with about 75\% of the front-end modules). In the fourth CRUZET
exercise, in August 2008, the complete silicon pixel tracker was introduced, along with
the endcaps of the ECAL.
In addition to the operational experience of the exercises and the ability
to address more subtle detector performance issues with larger event
samples, the data were critical for deriving zero-field alignment
constants for
the inner tracking systems. Several detector studies using CRAFT data
also made use of these CRUZET data samples. The total accumulated cosmic triggers
at zero field exceeded 300 million, including the triggers recorded
in September 2008 when the experiment was live for the first LHC beams.

These global runs  regularly exercised the full data flow from the
data acquisition system
at the experimental site to the reconstruction facility at the
CERN IT centre (called the Tier-0 centre), followed by the
subsequent transfer of the reconstructed data to all seven of the CMS
Tier-1 centres and to some selected Tier-2 centres \cite{Workflow}.



\subsection{Final Closing of CMS}

\begin{figure}
\begin{center}
\includegraphics[width = 400pt]{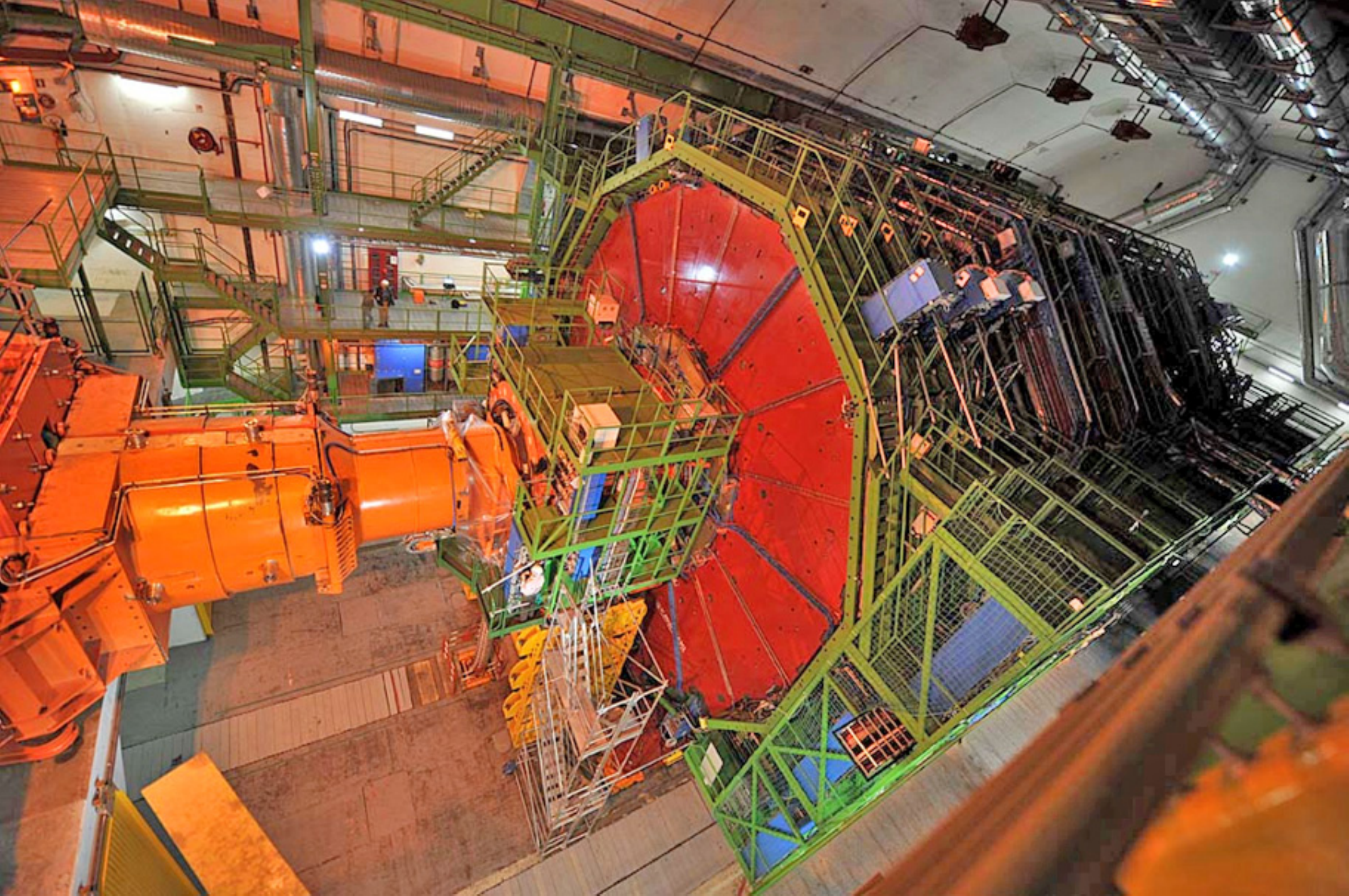}     
\caption{
The CMS experiment in its final, closed configuration in the
underground experimental cavern.
\label{fig:CMSClosed} }
\end{center}
\end{figure}

The final
sequence of closing the steel yoke and preparing CMS for collisions
was completed on August 25, 2008 (see
Fig. \ref{fig:CMSClosed}).
This was followed by several tests of the solenoid in the underground cavern
for the first time, up to a field
 of $3\,$T, as described further in Section~\ref{sec:magnet}.
The final test at the operating field
was
postponed until CRAFT due to the imminent arrival of beam at the beginning of
September and the necessity of keeping the solenoid off for the initial
commissioning phase of the LHC.

\subsection{LHC Beam Operations in 2008}

The CMS experiment was operational and recorded triggers associated
with activity from the first LHC beams in September 2008. This
activity included single shots of the beam onto a collimator $150\,$m
upstream of CMS, which yielded sprays (so-called ``beam splashes'')
containing $\mathcal{O}(10^5)$
muons crossing the cavern synchronously, and
beam-halo particles associated with the first captured orbits of the
beam on September 10 and 11.

The configuration of the experiment for LHC beam operations was nearly
the same as
that for CRAFT. The exceptions were that the silicon
pixel and strip tracking systems were powered off for safety reasons,
time delays in the readout electronics between the top and bottom halves of the experiment were removed,  and the Level-1 trigger menu
was set for synchronous beam triggers.

The first ``beam splash'' events were used to synchronize the beam triggers,
including those from the RF beam pick-ups, the beam scintillation
counters surrounding the beam pipe, the forward hadron calorimeters,
and the CSC muon system.
The diamond beam condition monitors were also commissioned with beam,
providing online diagnostics of the beam timing, bunch structure, and
beam-halo.
The data collected from the ``beam splash'' events also proved useful for
adjusting the inter-channel timing of the ECAL \cite{ECALTiming} and HCAL \cite{HCALTiming}
readout channels, as the
synchronous wave of crossing muons has a characteristic time-of-flight
signature.

In total, CMS recorded nearly 1 million beam-halo triggered events during the 2008 beam operations.

\section{Experiment Setup for CRAFT}
\label{sec:setup}

\subsection{Detector Components}

All installed detector systems were available for testing during
CRAFT.
The silicon pixel tracker and the ECAL endcaps were the last major systems
to be installed and thus were still being commissioned and tuned even
after the start of CRAFT.
Furthermore, the commissioning of the RPC system had been delayed by the
late delivery of power supplies; and by the time of CRAFT the RPC endcap disks
were not yet commissioned for operation.

\subsection{Trigger and Data Acquisition}

The typical Level-1 trigger rate during CRAFT was $600\,$Hz.
This rate is well below the $100\,$kHz
design limit for the central data acquisition (DAQ) system  and is composed of about
$300\,$Hz of cosmic triggers using all three muon systems, $200\,$Hz
of low threshold triggers
from the calorimeters, and $100\,$Hz of calibration triggers used to
pulse the front-end electronics or to illuminate the optical readout paths
of the calorimeters.
The cosmic muon triggers were more permissive than what would be
used during collisions, with only loose requirements for the
muon to point to the interaction region of the experiment. The rate of
triggered cosmic muons crossing the silicon strip tracker region was
 $\mathcal{O}(10)\,$Hz. As CMS is located $100\,$m below the surface
of the Earth, the cosmic muon rate relative to that at the surface is
suppressed by approximately
two orders of magnitude. The time-of-flight of cosmic muons to cross
from the top to the bottom of the experiment was accounted for
by introducing coarse delays of the muon trigger signals in the top half
such that they are in rough coincidence with the bottom
half (a two bunch crossing difference for the barrel, and one for the
endcaps, where one bunch crossing corresponds to $25\,$ns).
The calorimeter triggers were configured with low thresholds selecting mostly
detector noise.
Further details on the configuration and performance of the
Level-1 trigger during CRAFT can be found in Ref. \cite{CRAFTTrigger}.

The event filter farm was configured into four
``slices''.
There were 275
builder units that assembled the data fragments into complete
events, and 825 filter units for HLT processing.  The average size of an event built by the
DAQ during CRAFT is about $700\,$kB.
The HLT primarily applied only
pass-through triggers with no filtering, in order to efficiently
record cosmic ray events selected by the Level-1 trigger, although
additional complex filters were phased in for physics selection, alignment
and calibration of the detector, and diagnostics \cite{CRAFTHLT}.
At the end of CRAFT, a DAQ
configuration with eight slices and nearly 4500 filter units was
successfully tested.

\section{Operations}
\label{sec:operations}

\subsection{Control Room Operations and Tools}

Control room operations for CMS, as executed during CRAFT, are
carried out by a five-person
central shift crew
responsible for the global data-taking
of the experiment, and 10--14 subsystem shifters
responsible for the detailed monitoring and control of specific
detector systems during this commissioning period. These operations
are in general conducted continuously, necessitating three shift crews daily.

The central shift crew is composed of a shift leader, a
detector safety and operation shifter, a DAQ operator, a trigger
control operator, and a data quality monitoring (DQM) shifter. The shift
leader is responsible for the overall safety of the experiment and
personnel, and for the implementation of the run plan as set by the run
coordinators.

The DAQ operator issues the sequence of commands for initializing the
detector readout electronics and controls the data-taking state
via the run control system.
Several displays are devoted to monitoring the state of the DAQ
system and for detailed diagnostics.
The trigger operator is
responsible for the monitoring of the Level-1 trigger
system and for configuring
the system with the desired settings (the participating trigger
components, delays, etc.).

The central DQM shifter in the control room uses the CMS-wide DQM
services \cite{Workflow}, monitoring event data to assess the
quality of the data collected during a run. The DQM shifter is
responsible for certifying runs for data analysis purposes. This
information is entered into a ``run registry'' database (which also
contains configuration information about the run), and forms the first
step in a chain that assigns quality flags on a subsystem-by-subsystem
basis. The information was used to select event samples,
such as those used in the studies
of the detector performance during CRAFT. The control
room DQM shifter is assisted by another shifter located at one of the
remote operations centres (Fermilab and DESY).

Additional information available to the central and detector shifters
to assess the detector status includes the notification of any alarm
triggered by the detector safety system, such as cooling or
power failures, and monitoring
data from the Detector Control System (DCS) such as
temperature readings from the detector and electronic components, the
status of the cooling plants, the
status of the high and low voltage power, etc.
The DCS system is accessible from within the online network, and
graphical interfaces have been developed for use in the control room.
As the detector status data are stored in a database, a
set of software tools, known as Web-Based Monitoring (WBM), was also
designed to extract and display the information inside and outside the
control room network.
Both real-time status data and
historical displays are provided.
Example WBM applications are: Run Summary
(detailed information about the runs taken), DCS information (current condition
and past history of each subdetector component), magnet variables, and trigger rates.
These WBM applications were used extensively during the CRAFT
data-taking operation as well as during the subsequent analysis stage
to understand the data-taking configurations and conditions.

\subsection{Magnet Operations}
\label{sec:magnet}


The operating current for the solenoid was set to $18\,160\,$A
($B=3.8\,$T), although the magnet has been successfully tested to its
design current of $19\,140\,$A ($B=4.0\,$T) as noted in
Section~\ref{sec:mtcc}. This choice for the initial operating phase of the
experiment was made to have an additional safety
margin, with little impact on physics measurements,
in view of the long period of operation that
is expected to exceed 10
years.

The magnet operated successfully for the duration of CRAFT.
Nominal performance was a\-chieved in the control racks, safety
system, cryogenic system, and passive protection system.
Apart from a ramp-down to allow access to the experimental cavern
during the LHC inauguration,
the only interruptions of
the magnet were due to
water cooling interlocks caused by an
incorrectly adjusted leak-detection threshold.

\begin{figure}
\begin{center}
\includegraphics[width = 300pt]{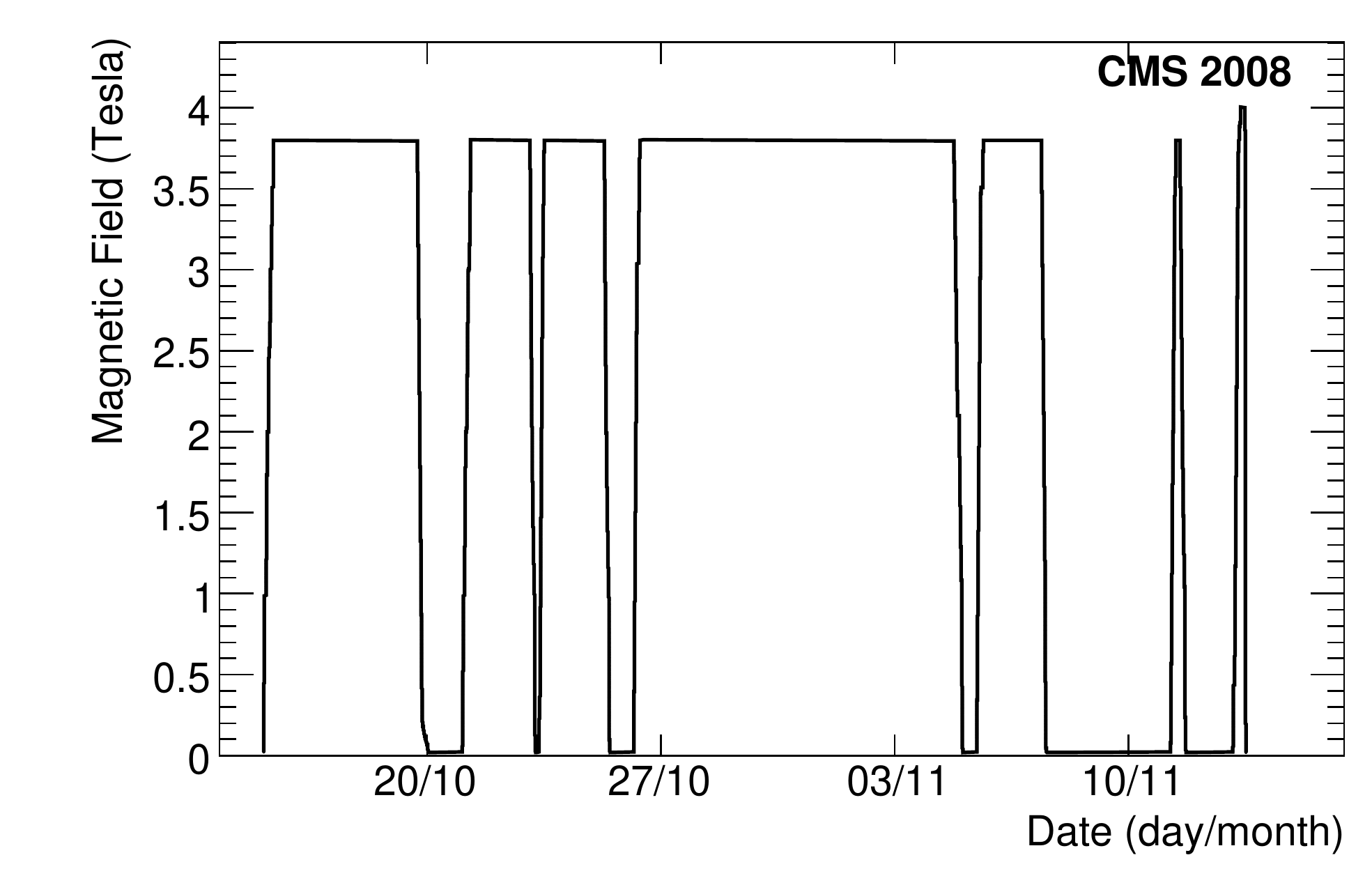}     
\caption{History of the magnitude of the central magnetic field during
CRAFT. The ramp-down on October 20 was needed to allow open access to
the experimental cavern. The last two ramp-ups in November were
further tests of the magnet after CRAFT data-taking.
\label{fig:history} }
\end{center}
\end{figure}

The magnet was at its operating current
for the CRAFT
exercise for a total
of 19 days, between October 16 and November 8, 2008,  as illustrated in
Fig.~\ref{fig:history}.
Ramping tests indicate a nominal time of 220
minutes for magnet rampings (up or down), keeping the magnet at a temperature
of $4.5\,$K.
Distortions of the yoke during
ramps were measured by the muon alignment systems (Section~\ref{sec:align}).

Shortly after the CRAFT data-taking exercise at $3.8\,$T ended,
a final ramp was made to $4.0\,$T on November
14 to ensure
the magnet stability margin. After bringing the field back to $3.8\,$T,
a fast discharge was triggered, which took only 10 minutes.
The final average temperature of the
magnet after a fast discharge is $66\,$K, requiring at least three days to
reestablish the operating temperature.

\subsection{Infrastructure}

The infrastructure and services met the demands of
running the experiment continuously for one month, although the exercise
indicated areas needing further improvement.
No particular problem or malfunction of the
electrical and gas distribution systems for the experiment occurred
during CRAFT. Likewise,
the nitrogen inerting and dry air systems,
intended to prevent fire and guarantee a dry atmosphere in humidity
sensitive detectors, operated stably.
The detector safety and
control systems performed as expected, but further functionality
and testing took place after CRAFT.

Several cooling failures did occur, and this resulted in the
shutdown of some equipment during CRAFT.
One circulating pump failed due to a faulty installation.
Leaks were detected on the barrel yoke circuit for wheels YB$-2$ and YB$-1$.
As noted earlier, the leak detection system on one of the cooling circuits fired a few
times resulting in three separate
automatic slow dumps of the magnet (leading to at least 8 hours loss at
$3.8\,$T). The threshold has been subsequently raised to make the system
more robust.

The cooling plants for the silicon strip tracker suffered a few trips,
leading to about 6 hours down-time during CRAFT. The leak rate of the
system was also higher than expected. In the shutdown after CRAFT, the
cooling plants and piping were significantly refurbished to
eliminate these leaks.

In total, about 70 hours of downtime were caused by general infrastructure
related incidents, about 10\% of the duration of CRAFT. This
time was dominated by the downtime of the magnet.

\subsection{Detector Operations}

\nocite{StripPerf}
\nocite{PixelPerf}
\nocite{ECALPerf}
\nocite{HCALPerf}
\nocite{DTLocalTrig}
\nocite{RPCPerf}
\nocite{CSCLocalReco}
The operational performance of the detector subsystems during CRAFT
is reported in detail in Refs.
\cite{StripPerf}--\cite{CSCLocalReco}. All detector systems
functioned as intended in the $3.8\,$T magnetic field.
Here we summarize the principal
operations conducted during CRAFT and the main observations.

\subsubsection{Tracker}

The silicon strip tracker \cite{StripPerf} was live 95\% of the running time during
CRAFT, with 98\% of the channels active.
Signals were collected via a $50\,$ns CR-RC shaper, sampled and stored
in an analog pipeline by the APV25 front-end chip \cite{APV}.
The APV25 chip also contains a deconvolution circuit, not used during CRAFT,
to reduce the signal width (but increasing the noise) that can be
switched on at high LHC luminosity, when pile-up becomes an issue.
The tracker readout was synchronized to triggers delivered by the muon
detectors.
A few issues not identified during the previous commissioning of the
detector, such as some swapped cables and incorrect fibre length
assumptions used in the latency calculations, were quickly identified
by offline analysis of the cosmic data and corrected either during
operation or the subsequent shutdown.
Data were zero suppressed during the entire
exercise. The signal-to-noise ratio was excellent, ranging
from 25 to more than 30 depending on the partition,
after final synchronization adjustments.

The silicon pixel tracker \cite{PixelPerf} was live 97\% of the running time.
In the barrel pixel system, 99\% of the channels were
active, whereas in the forward pixel system 94\% were active. The 6\%
inactive channels in the latter were mostly
due to identified shorts in the power supply cables, which were
repaired during the subsequent shutdown.
Zero suppression was performed on the detector, with conservative
thresholds of about 3700 electrons chosen to ensure stable and efficient operations during CRAFT.
The pixels were mostly immune to noise, with a noisy channel fraction
of less than $0.5\times 10^{-5}$.   The mean noise from the front-end
readout chips in the barrel and forward detectors is 141 and 85
electrons, respectively, well below the operating thresholds.

\subsubsection{ECAL}

For the ECAL \cite{ECALPerf}, the fraction of channels that were operational during CRAFT was
98.5\% in the barrel and 99.5\% in the endcaps. A large
part of the barrel inefficiency was due to a cut power cable that
has since been repaired.
In the barrel, approximately 60\% of the dataset was recorded with
nominal front-end electronics gain while the other 40\% was recorded
with the gain increased by a factor of 4, to enhance the muon signal.

The response of the ECAL electronics, both in the barrel and in the endcap,
was monitored using  pulse injections at the preamplifier, showing no
significant changes due to the magnetic field.
Noise levels were generally consistent with the values measured during
construction, aside from a small increase for 1/4 of the
barrel that is believed to be low frequency pickup noise associated with the
operation of other CMS subdetectors and
that is mostly filtered by the amplitude reconstruction
algorithm.


The temperature of the ECAL is required to be stable at the level of
$0.05\,^\circ$C in the barrel and $0.1\,^\circ$C in the endcaps in order
to ensure that temperature fluctuations remain a
negligible contribution to the  energy resolution (both crystal light
yield and barrel front-end gain are temperature dependent). The
stability provided by the temperature control system during
CRAFT was measured to be
about $0.01\,^\circ$C for the barrel and  $0.05\,^\circ$C in the endcaps,
with almost all measurements   better than the required
specifications.

The ECAL barrel  high voltage should also be kept stable at the level
of a few tens
of mV due to the strong gain dependence of the photodetector on the
absolute voltage
value (3\% per volt).
The average fluctuation of the high voltage is measured to be $2.1\,$mV (RMS),
with all channels within $10\,$mV.

A laser monitoring system
is critical for maintaining the stability of the constant term
of the energy resolution at high luminosities.
Its main purpose is to
measure transparency changes for each crystal at the 0.2\% level, with
one measurement every 20--30 minutes. During CRAFT, a total of
approximately 500 sequences of laser monitoring data were taken, with
each sequence injecting 600 laser pulses per crystal. These data were
collected using a calibration trigger issued in the LHC abort gap at
a rate of typically $100\,$Hz. The measured stability was better than the
0.2\% requirement for 99.9\% of the barrel and 99\% of the endcap crystals.




To stabilise the response of the endcap VPTs, which have a rate-dependent gain (5--20\% variation in
the absence of
magnetic field, and significantly reduced at $3.8\,$T \cite{ECALPerf}), an LED pulsing system was
designed to continuously pulse the VPTs with a rate of at least
$100\,$Hz. This system was successfully tested during CRAFT on a small subset
of channels, and LED data were
acquired in different configurations.

\subsubsection{HCAL}

HCAL participated in the CRAFT data-taking with all components: barrel
(HB), endcap (HE),
forward (HF) and outer (HO) calorimeters \cite{HCALPerf}.
The fraction of non-operational channels overall for HB, HE and HF was 0.7\%
(0.5\% due to noisy HPDs in HB and HE, and 0.2\% to
electronics failures), while for HO it was about 4.5\%
(3.3\% due to noisy HPDs and 1.2\% to electronics)
at the start of CRAFT and increased to 13\% due to HPD failures as noted below.


As found during the MTCC exercise, the HPD noise rate
depends on the magnetic field. Therefore,
the behaviour of all HPDs was carefully monitored during CRAFT to
identify those HPDs that failed, or were likely to fail, at $3.8\,$T in
order to target them for replacement.
Noise data from HB and HE were collected using a trigger
with a threshold of $50\,$fC that is approximately equivalent to $10\,$GeV
of energy.
Based on individual HPD discharge rates, the high voltage to four
HPDs on HB and HE (out of 288 total)
was reduced during CRAFT from 7.5 to $6.0\,$kV (which lowers the gain by
approximately 30\%), in addition to two HPDs that were completely turned off.
At $3.8\,$T the resulting measured trigger rate
from 286 HPDs in HB and HE was  approximately
$170\,$Hz, which can be compared to the rate at zero field of about 130$\,$Hz.
The HPDs of HB and HE showed no signs of increased noise
rates during CRAFT.


The HO HPDs servicing the central wheel (YB$0$) operate in a fringe
field of $0.02\,$T (when the central field of the magnet is
$3.8\,$T) while those on  the outer wheels (YB$\pm 1$, YB$\pm 2$)
experience a magnetic
field above $0.2\,$T. While no HPDs on YB$0$ showed any significant
discharge rate,
it was expected from the MTCC experience that several HPDs on the
outer wheels would, and this was observed.
Twenty HPDs installed on the outer wheels at the
start of CRAFT showed
significant increase in the  noise rates at $3.8\,$T with respect to
the noise rates at $0\,$T.  The high voltage
was turned off for the four HPDs with the highest noise rate increases, and
was lowered to $7\,$kV for the others.
The HO HPDs located on the outer wheels
showed clear signs of increased discharge rates during CRAFT.
As the number of discharging HPDs continued to increase,
the high voltage on all HPDs servicing the outer
wheels was further lowered to $6.5\,$kV midway into CRAFT.
By the end of the run, a total of 14 HPDs were turned off out of
132 for all of HO.

During the winter 2008/09 shutdown, after CRAFT, the
problematic channels were fixed, including replacing HPDs with anomalously
high noise rates or low gains. A total of 19 HPDs were replaced in HB and HE,
and another 19 in the HO outer wheels.
As of November 2009, after additional tests with the magnetic field,
all of the HB, HE, and HF channels were operational while the number of
non-operational channels in HO was at the level of 4\%.

\subsubsection{Muon Detectors}


The DT system had all 250 chambers installed, commissioned  and
equipped with readout and trigger electronics for CRAFT, and 98.8\%
of the channels were active.
The DT
trigger was operated in a configuration requiring the coincidence of at
least two nearby chambers
without
requiring that the muon trajectory points to the nominal interaction
point.  The cosmic trigger rate for this configuration was stable at
about $240\,$Hz. The
performance of the DT trigger is
described in more detail in Ref. \cite{DTLocalTrig}.

The DT system demonstrated high reliability and stability.  Some
noise was observed sporadically during the period when the field was
on. In particular, synchronous noise produced huge events, with high
occupancy in the detectors comprising the wheels on one side of the
experiment, at the level of 0.1--1$\,$Hz. Noise sources are under investigation.


The RPC system participated in CRAFT with the entire barrel and a small
fraction of the
endcaps \cite{RPCPerf},  which at the time were at an early stage of commissioning.
For the barrel part,  the CRAFT  operation  was important to ascertain the
system stability,
debug the hardware,  synchronize the electronics, and ultimately obtain
a preliminary measurement of the
detector performance.
About 99\% of the barrel electronic channels were active during the data taking,
while the remaining 1\% were masked  due to a high counting rate. The
average cosmic muon RPC trigger rate was about $140\,$Hz for the barrel,
largely coincident with the
DT triggers, with some spikes in rate as noted below.

The main RPC monitoring tasks ran smoothly during the entire period, which allowed
a careful
analysis of system stability. The average current drawn by the  barrel
chambers (each having a $12\,$m$^2$
single gap surface) was stable below $1.5\,\mu$A, with  very few cases
above $3\,\mu$A.
A  study of the chamber
efficiency as a function of the operating voltage
was possible for about 70\% of the barrel chambers, giving  a preliminary
estimate of the  average intrinsic detector
efficiency of  about 90\%.
This study also indicated a few hardware failures and cable map errors
that were later fixed.

Sporadic RPC trigger spikes related  to noise   pick-up
 from external sources were also detected.
The sensitivity of the system  to  these
sources is under investigation. However, preliminary studies have
demonstrated that
the trigger rate is almost unaffected   when the standard trigger algorithm
for LHC collisions is  applied.


The CSC system operated with more than 96\% of the readout channels active and
for about 80\% of the CRAFT running period \cite{CSCLocalReco}.
The rate of trigger primitive segments  in the CSCs from cosmic-ray
muons underground was about $60\,$Hz,
distributed over the two endcaps.
The CSC trigger
was configured to pass each of these trigger segments, without a
coincidence requirement,
as muon candidates to the Level-1 Global Trigger.

The long running period under stable conditions provided by
CRAFT exposed a few issues that had not yet been encountered during the
commissioning of the CSCs.  These effects include a very low
corruption rate of the non-volatile memories used to program some
FPGAs distributed in the system, and unstable communication with the
low voltage power supplies.  Actions taken after CRAFT to address
these issues included periodic refreshing of the memories and
a replacement of the control signal cables to the power supplies, respectively.

\subsubsection{Muon Alignment System}
\label{sec:align}

The complete muon alignment system was tested during
CRAFT \cite{MuonHWAlign}. It is organized into three main
components: two local systems to monitor the relative positions of the DT and CSC
muon detectors separately, and a ``link system'' that relates the muon chambers and central tracker
and allows  simultaneous monitoring of the barrel and endcap.
All components are designed to
provide continuous monitoring of the muon chambers in the entire magnetic field
range between $0\,$T and $4\,$T. The acquisition of data from these systems
is separate from the central DAQ used to collect normal event data.

Each DT chamber is equipped with LEDs as light sources, and about 600
video cameras mount\-ed on rigid carbon-fibre structures observe the
motions of the chambers. During the CRAFT data taking period, as well
as the periods just before and after, over
100 measurement cycles were  recorded.
Compression of the barrel wheels in $z$ with the
magnet at $3.8\,$T was observed at the level of 1--2$\,$mm depending
on azimuth, and with an
uncertainty of $0.3\,$mm, in agreement with previous measurements made
during  MTCC.

The acquisition of endcap
alignment data  was robust, and 99.4\%
of the sensors were operational. The sensors monitor displacements
from reference laser beams across each endcap disk, and provide
positioning information on 1/6 of the endcap chambers.
This allows adequate monitoring of the yoke disk deformations due to
strong magnetic forces.
The measurements with the magnet at $3.8\,$T indicate
deformations of all disks towards the interaction point by about
10--12$\,$mm for chambers close to the beam line and by about $5\,$mm for
chambers further away.  These results are consistent with earlier
MTCC measurements.

The link system comprises amorphous silicon position detectors
placed around the muon spectrometer and connected by laser lines.
The complete system was implemented for CRAFT, and
98\% operational efficiency was obtained. Unfortunately, the closing
of the YE$-1$ disk
outside of its tolerance created a conflict with some of the alignment
components, making the laser system for this part of the detector effectively
unusable during CRAFT. The disk could not be repositioned given the
limited remaining time to prepare CMS for LHC beams in 2008.
Consequently, full reconstruction of the link system data was possible
only for the $+z$ side of the
detector. During CRAFT, data were recorded at stable
$0\,$T and $3.8\,$T field values, as well as during magnet
ramps.
Information from the link system was used to align CSCs on YE$\pm1$.
Both the endcap alignment and the link system detected disk bending, and CSC tilts
were measured at full field.


\subsection{Data Operations}

\begin{figure}
\begin{center}
\includegraphics[width = 400pt]{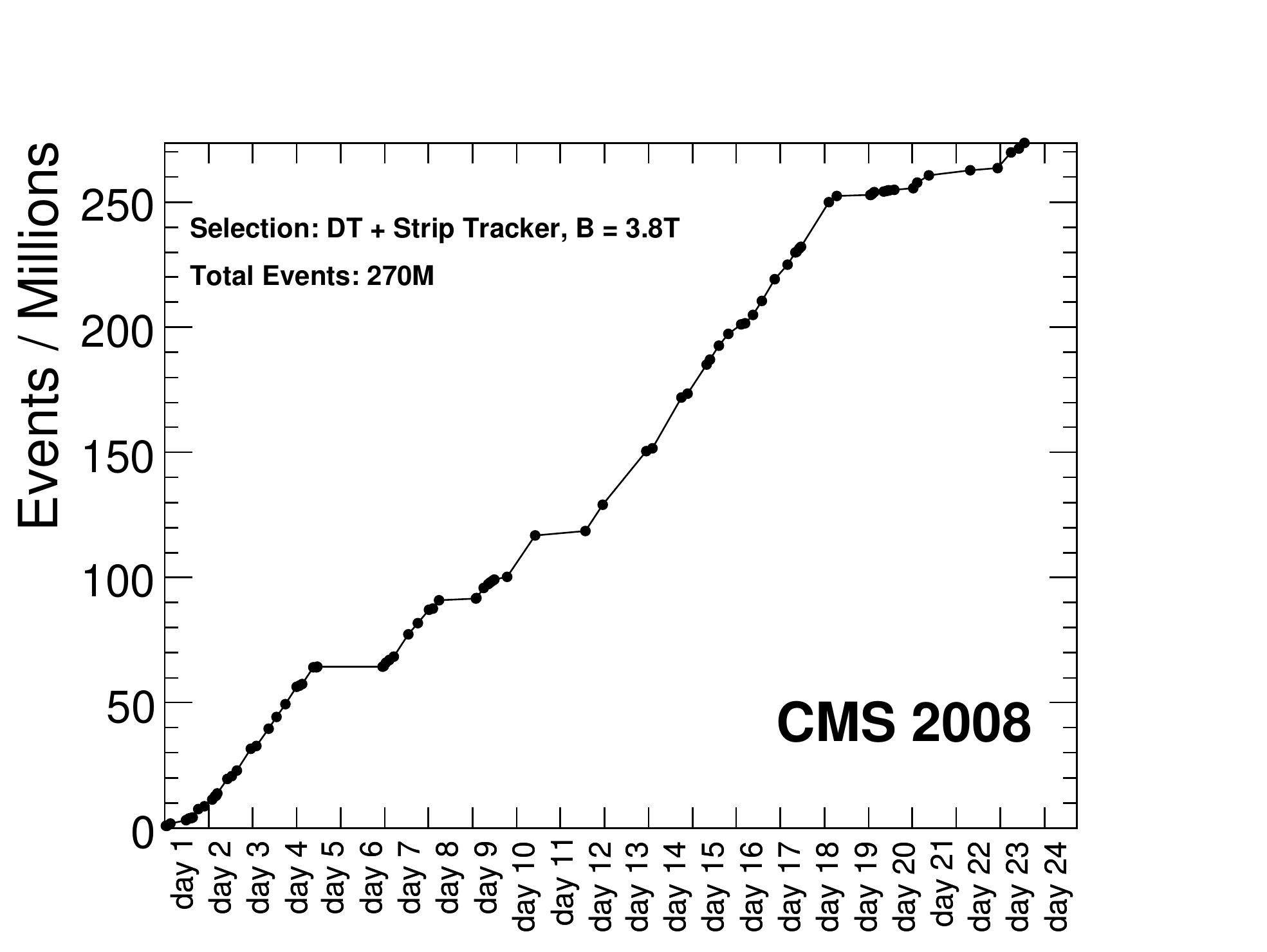}     
\caption{The accumulated number of good (see text) cosmic ray triggered
events with the magnet at $3.8\,$T as a function of days into CRAFT,
beginning October 16, 2008.
\label{fig:stats} }
\end{center}
\end{figure}

The average data-taking run length during CRAFT was slightly more than 3 hours,
and four runs exceeded 15 hours without interruption (one
run exceeded 24 hours). Reasons for stopping a run included the desire to
change the detector configuration, a hardware-related issue with a
detector subsystem (e.g. loss of
power), some part of  the readout electronics entering an
irrecoverable busy state, or other irrecoverable DAQ system errors. Aside from the 10\% downtime due to
infrastructure related problems, the typical data collection efficiency of CMS
was about 70\% during CRAFT, including  periods used to
conduct detector calibrations, pedestal runs, and diagnostics
to further advance the commissioning of the experiment.
The breakdown of the total number of collected
events passing quality criteria for the detector systems or
combination of systems, but
not necessarily with a cosmic muon within its fiducial volume,
is listed in Table~\ref{tab:allevents}.
Figure~\ref{fig:stats} shows the
accumulated number of cosmic ray triggered events as a function of
time with the magnet at
its operating central field of $3.8\,$T, where
the minimal configuration of the
silicon strip tracker and the DT muon system delivering data
certified for further offline analysis was required.
It was not required to keep the other systems in the configuration.
A total of 270 million such
events were collected. The effective change in slope after day 18 is
principally due to downtime to further improve the synchronization of
the silicon strip tracker and an unplanned ramp-down of the magnet.

\begin{table}
\centering
\caption{The number of cosmic ray triggered events collected
during CRAFT
with the magnetic field at its operating axial value of $3.8\,$T
and with the listed detector system (or combination of systems)
operating nominally and passing offline quality criteria.
The minimum configuration required for data-taking was at least one
muon system and the strip tracker. The other subdetectors were allowed
to go out of data-taking for tests.
}
\label{tab:allevents}
\vskip 10pt
\begin{tabular}{l|r}
\hline
Detector & Events (millions)  \\
\hline
Pixel Tracker & 290  \\
Strip Tracker & 270  \\
ECAL & 230  \\
HCAL & 290 \\
RPC & 270  \\
CSC & 275  \\
DT & 310  \\
DT+Strip & 270  \\
All & 130  \\
\hline
\end{tabular}
\end{table}

Data were promptly reconstructed at the Tier-0
computing centre at CERN to create high-level physics objects
with a job latency of about 8 hours, but with a broad distribution.
These data were transferred to
the CMS Analysis Facility (CAF) at the CERN Meyrin site and to several
Tier-1 and Tier-2 centres worldwide  for
prompt analysis by teams of physicists. The average export rate from
the Tier-0 centre to the Tier-1 centres during CRAFT was $240\,$MB/s, and
the total volume transferred was about $600\,$TB.  Data quality monitoring of the
Tier-0 reconstruction output in addition to the standard online
monitoring at the control room was provided.
Specialized data selections for detector calibration and alignment purposes
also were created from the Tier-0 processing, and they were processed on
the CAF to update the alignment and calibration constants.
Several reprocessings of the CRAFT datasets took place at the Tier-1 centres
using the refined calibration and alignment constants after the CRAFT
data taking period.
For ease of offline analyses, several primary datasets were produced that were
derived from
dedicated HLT filters. Some datasets were further reduced,
selecting for example events enriched in the
fraction of cosmic muons
pointing to the inner tracking region of the experiment.
Further details on the
offline processing workflows applied to the CRAFT data are described
in Ref. \cite{Workflow}.

\subsection{Lessons}

The extended running period of CRAFT led to the identification of some
areas needing attention for future operations. As noted in a few instances
already, some repairs or replacements of detector subsystem components
were required, such as the repair of several power cables to the pixel
tracker, the replacement of some HPDs for the HCAL, and the
replacement of some muon electronics on the detector. 
The cooling plants for the silicon tracking systems were refurbished
to eliminate leaks in the piping, and improved water 
leak detection was added to the barrel yoke. 
Based partly on the experience of CRAFT, the complexity of the HLT
menu to be used for initial LHC operations was reduced 
and some algorithms were improved for better CPU and memory usage
performance. 

In order to measure the efficiency of data-taking automatically and to
systematically track the most significant problems, a tool
was developed after CRAFT to 
monitor the down-times during data-taking and the general reasons for
each instance (as selected by the shift leader). In an effort to
improve the efficiency, further centralization of operations
and additional alarm capability were added to the detector control and
safety systems.
More stringent change-control policies on hardware interventions, firmware
revisions, and software updates were also enacted during subsequent 
global-taking periods in order to further limit any unannounced changes.  

Many CRAFT detector analyses successfully used the computing
facilities of the CAF, but to the extent that it became clear
afterward that the disk and CPU usage policies needed refinement.
One unexpected outcome from the CRAFT analyses was the identification
of a problem in the initial
calculation of the magnetic field in the steel yoke for the barrel
that became evident when studying cosmic muons recorded during
CRAFT (see Section~\ref{sec:performance}).

\section{Detector Performance Studies}
\label{sec:performance}

\begin{figure}
\begin{center}
\includegraphics[width = 400pt]{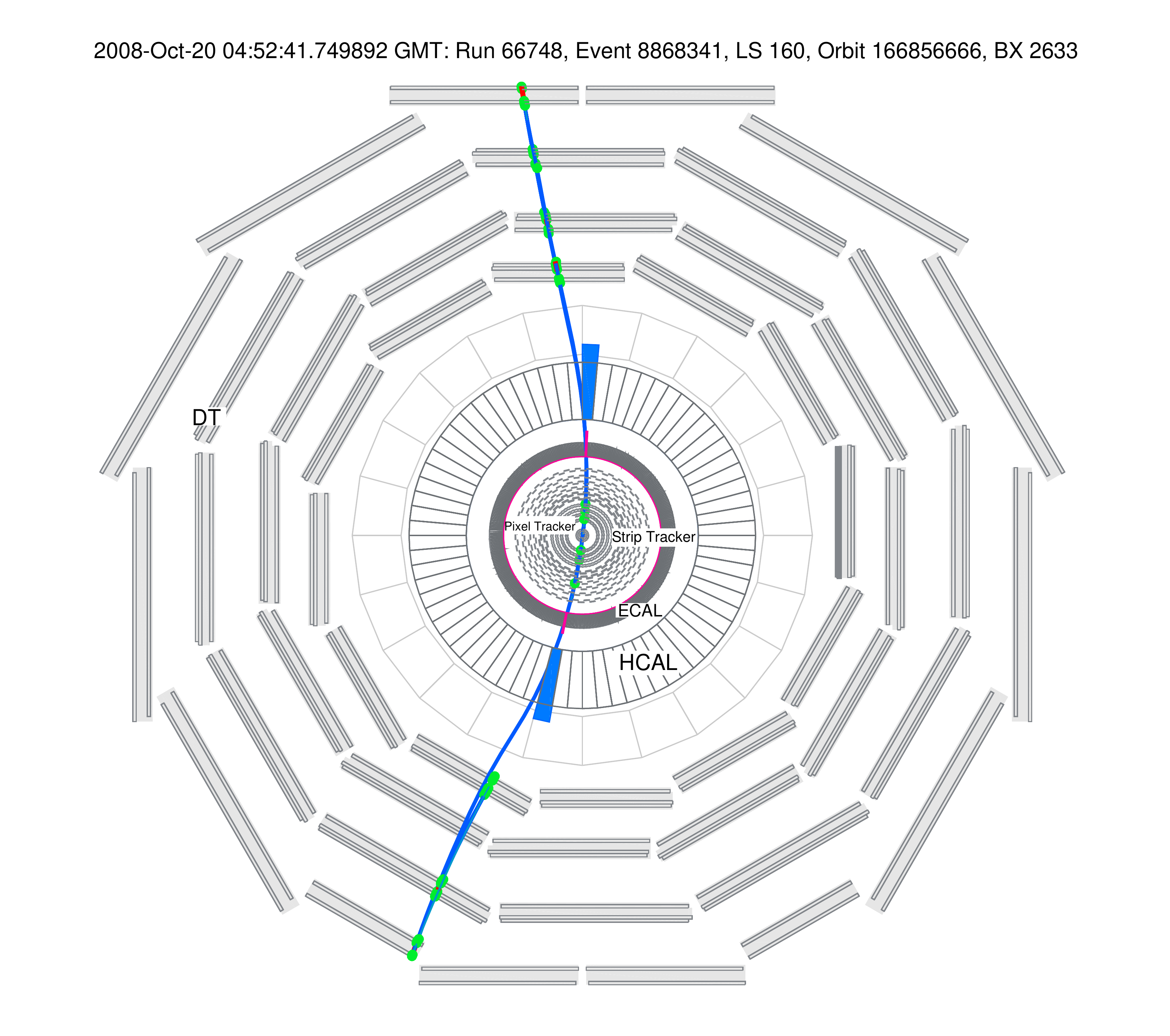}     
\caption{Display of a cosmic muon recorded during CRAFT which enters
and exits through the DT muon system, leaves
measurable minimum ionizing deposits in the HCAL and ECAL,
and crosses the silicon strip and pixel tracking systems.
Reconstruction of the trajectory is also indicated.
\label{fig:EVD} }
\end{center}
\end{figure}

The data collected during CRAFT, with CMS in its final
configuration and the magnet energized, facilitated  a wide range of
analyses on the performance of the CMS subdetectors, the magnitude of the
magnetic field in the return yoke, as well as the
calibration and alignment of sensors in preparation for physics
measurements.
Figure \ref{fig:EVD} shows a transverse view of the CMS
experiment with the calorimeter energy deposits and
tracking hits indicated from one cosmic muon traversing the central region during
CRAFT.
It also shows the results of the muon trajectory reconstruction.

Alignment of the silicon strip and pixel sensor modules
was improved significantly from initial survey measurements
by applying
sophisticated track-based alignment techniques to the data recorded from
approximately 3.2 million tracks selected to cross the sensitive
tracking region
(with $110\,000$ tracks having at least one
pixel hit). The precision achieved for the positions of the detector
modules with respect to particle trajectories, derived from the
distribution of the
median of the cosmic muon track residuals, is
3--4 $\mu$m in the barrel and 3--14 $\mu$m in the endcaps for the
coordinate in the bending plane \cite{TrackerAlign}.
Other silicon tracking measurements \cite{StripPerf}
performed with the CRAFT
data include calibration of the absolute energy loss in silicon strip sensors,
Lorentz angle measurements, hit efficiencies and position resolutions, track
reconstruction efficiencies, and track parameter resolutions.
The efficiency of reconstructing cosmic ray tracks, for example, is greater than 99\%
for muons passing completely through the detector and close to the
beam line.

Track-based alignment techniques using cosmic muons were also applied to align the
DT muon detectors in the barrel region of the experiment. An alignment
precision of better than 700 $\mu$m was achieved along the higher precision
coordinate direction (approximately $\phi$) for the first three DT
stations as estimated by a cross-check
of  extrapolating muon segments from one detector to
the next \cite{MuonTBA}.
A local alignment precision of $270\,\mu$m was achieved within
each ring of CSCs
using LHC beam-halo muons recorded during beam operations in 2008.

Studies of the resolution and efficiency of hit reconstruction
in the DT \cite{DTLocalReco} and CSC \cite{CSCLocalReco} muon
chambers were
performed. The resolution of a single reconstructed hit in a DT
layer is of the order of $300\,\mu$m, and the efficiency of
reconstructing high quality local track segments built from several
layers in the DT chambers is approximately
99\%. Likewise, the position resolution of local track segments in
a CSC is of the order of $200\,\mu$m ($50\,\mu$m for the highest $|\eta|$
chambers on YE$\pm 1$). The  performance of the muon track
reconstruction algorithms on CRAFT data was
studied \cite{MuonReco} using the muon system alone and using the
muon system combined with the inner tracker.
The requirement on the distance of closest approach to the
beam line must be relaxed relative to that used for muons from collisions.
The resolution of the track parameters can be determined by
splitting a single cosmic ray signal into upper and
lower tracks and comparing the results of the separate fits.
For example, the $p_{\rm T}$
resolution for tracks passing close to the beam line with a sufficient
number of hits in the silicon pixel and strip tracking detectors
is measured to be of the order of 1\% for
$p_{\rm T}=10\,$GeV/$c$, increasing to 8\% for $p_{\rm T}$
of about $500\,$GeV/$c$.
The latter is limited by the accuracy of the alignment of the inner tracker
and the muon system, and should improve to 5\% when perfectly aligned.

Measurements of the cosmic muon energy loss, ${\rm d}E/{\rm d}x$, in the ECAL and HCAL
barrel compartments validated the initial
calibration of
individual channels obtained prior to CRAFT (the endcap studies suffer
from small sample sizes).
In a study of 40\% of the ECAL barrel channels, the obtained spread in
detector response has an RMS of about 1.1\%, consistent with
the combined statistical and systematic uncertainties
\cite{ECALPerf}. Additionally, the measured ${\rm d}E/{\rm d}x$ as a
function of muon momentum agrees well with a first-principles
calculation \cite{ECALStopping}. For the HCAL
barrel and endcap, CRAFT confirmed the brightening of scintillators
with magnetic field
first measured during MTCC, which leads
to about a 9\% increase in response. The response to
cosmic muons recorded during CRAFT was used to adjust the
intercalibration constants of the barrel, and the residual RMS spread
is at the level of a few percent \cite{HCALPerf}. The absolute
response to cosmic muons with a momentum of $150\,$GeV/$c$ agrees well
with earlier test beam measurements.

The accuracy of the calculated magnetic field map in the barrel steel
yoke, used in muon reconstruction, was obtained by a
comparison of the muon bending measured by DT chambers with the
bending predicted by extrapolating the track parameters measured by
the inner tracking system (where the magnetic field is known very
precisely).
During CRAFT a discrepancy was noted, and this was later traced to
boundary conditions that had been set too restrictively in the field
map calculation. The analysis also suggested improving the treatment
of asymmetric features in the map. An updated
field map was produced based on these results. Residual differences
between data and the calculation are reduced to about 4.5\% in the
middle station of the barrel yoke and 8.5\% in the outer station, and are
corrected using the CRAFT  measurements \cite{BField}.

\section{Summary}
\label{sec:conclusion}

The CRAFT exercise in 2008 provided
invaluable experience in commissioning and operating the  CMS experiment and, from
analyses performed on the data recorded, in understanding
the performance of its detectors.
It represented a milestone in a global commissioning programme
marked by a series of global data-taking periods with progressively
larger fractions of the installed detectors participating, culminating
in all
installed systems read out in their entirety or nearly so.
Over the course of 23 days during
October and November 2008, the experiment collected 270 million  cosmic ray
triggered events with
the solenoid at its operating central field of
$3.8\,$T and with at least the silicon strip tracker and drift tube muon system
participating and operating at nominal conditions. These data were processed by the offline data handling
system,
and then analyzed by teams dedicated to the calibration, alignment,
and characterization of the detector subsystems.
The precision achieved of detector calibrations and alignments, as well as
operational experience of running the experiment for an extended
period of time, give confidence that the CMS experiment is ready
for LHC beam operations.

\section*{Acknowledgments}
We thank the technical and administrative staff at CERN and other
CMS Institutes, and acknowledge support from:
FMSR (Austria);
FNRS and FWO (Belgium);
CNPq, CAPES, FAPERJ, and FAPESP (Brazil);
MES (Bulgaria);
CERN;
CAS, MoST, and NSFC (China);
COLCIENCIAS (Colombia);
MSES (Croatia);
RPF (Cyprus);
Academy of Sciences and NICPB (Estonia);
Academy of Finland, ME, and HIP (Finland);
CEA and CNRS/IN2P3 (France);
BMBF, DFG, and HGF (Germany);
GSRT (Greece);
OTKA and NKTH (Hungary);
DAE and DST (India);
IPM (Iran);
SFI (Ireland);
INFN (Italy);
NRF (Korea);
LAS (Lithuania);
CINVESTAV, CONACYT, SEP, and UASLP-FAI (Mexico);
PAEC (Pakistan);
SCSR (Poland);
FCT (Portugal);
JINR (Armenia, Belarus, Georgia, Ukraine, Uzbekistan);
MST and MAE (Russia);
MSTDS (Serbia);
MICINN and CPAN (Spain);
Swiss Funding Agencies (Switzerland);
NSC (Taipei);
TUBITAK and TAEK (Turkey);
STFC (United Kingdom);
DOE and NSF (USA).
Individuals have received support
from the Marie-Curie IEF program (European Union); the Leventis
Foundation; the A. P. Sloan Foundation; and the Alexander von Humboldt
Foundation.

\bibliography{auto_generated}   
\clearpage
\appendix
\cleardoublepage\appendix\section{The CMS Collaboration \label{app:collab}}\begin{sloppypar}\hyphenpenalty=500\input{CFT-09-008-authorlist.tex}\end{sloppypar}
\end{document}

%% file: ptdr-definitions.tex
%
%
%

\newcommand {\etal}{\mbox{et al.}\xspace} 
\newcommand {\ie}{\mbox{i.e.}\xspace}     
\newcommand {\eg}{\mbox{e.g.}\xspace}     
\newcommand {\etc}{\mbox{etc.}\xspace}     
\newcommand {\vs}{\mbox{\sl vs.}\xspace}      
\newcommand {\mdash}{\ensuremath{\mathrm{-}}} 

\newcommand {\Lone}{Level-1\xspace} 
\newcommand {\Ltwo}{Level-2\xspace}
\newcommand {\Lthree}{Level-3\xspace}

\providecommand{\ACERMC} {\textsc{AcerMC}\xspace}
\providecommand{\ALPGEN} {{\textsc{alpgen}}\xspace}
\providecommand{\CHARYBDIS} {{\textsc{charybdis}}\xspace}
\providecommand{\CMKIN} {\textsc{cmkin}\xspace}
\providecommand{\CMSIM} {{\textsc{cmsim}}\xspace}
\providecommand{\CMSSW} {{\textsc{cmssw}}\xspace}
\providecommand{\COBRA} {{\textsc{cobra}}\xspace}
\providecommand{\COCOA} {{\textsc{cocoa}}\xspace}
\providecommand{\COMPHEP} {\textsc{CompHEP}\xspace}
\providecommand{\EVTGEN} {{\textsc{evtgen}}\xspace}
\providecommand{\FAMOS} {{\textsc{famos}}\xspace}
\providecommand{\GARCON} {\textsc{garcon}\xspace}
\providecommand{\GARFIELD} {{\textsc{garfield}}\xspace}
\providecommand{\GEANE} {{\textsc{geane}}\xspace}
\providecommand{\GEANTfour} {{\textsc{geant4}}\xspace}
\providecommand{\GEANTthree} {{\textsc{geant3}}\xspace}
\providecommand{\GEANT} {{\textsc{geant}}\xspace}
\providecommand{\HDECAY} {\textsc{hdecay}\xspace}
\providecommand{\HERWIG} {{\textsc{herwig}}\xspace}
\providecommand{\HIGLU} {{\textsc{higlu}}\xspace}
\providecommand{\HIJING} {{\textsc{hijing}}\xspace}
\providecommand{\IGUANA} {\textsc{iguana}\xspace}
\providecommand{\ISAJET} {{\textsc{isajet}}\xspace}
\providecommand{\ISAPYTHIA} {{\textsc{isapythia}}\xspace}
\providecommand{\ISASUGRA} {{\textsc{isasugra}}\xspace}
\providecommand{\ISASUSY} {{\textsc{isasusy}}\xspace}
\providecommand{\ISAWIG} {{\textsc{isawig}}\xspace}
\providecommand{\MADGRAPH} {\textsc{MadGraph}\xspace}
\providecommand{\MCATNLO} {\textsc{mc@nlo}\xspace}
\providecommand{\MCFM} {\textsc{mcfm}\xspace}
\providecommand{\MILLEPEDE} {{\textsc{millepede}}\xspace}
\providecommand{\ORCA} {{\textsc{orca}}\xspace}
\providecommand{\OSCAR} {{\textsc{oscar}}\xspace}
\providecommand{\PHOTOS} {\textsc{photos}\xspace}
\providecommand{\PROSPINO} {\textsc{prospino}\xspace}
\providecommand{\PYTHIA} {{\textsc{pythia}}\xspace}
\providecommand{\SHERPA} {{\textsc{sherpa}}\xspace}
\providecommand{\TAUOLA} {\textsc{tauola}\xspace}
\providecommand{\TOPREX} {\textsc{TopReX}\xspace}
\providecommand{\XDAQ} {{\textsc{xdaq}}\xspace}

\newcommand {\DZERO}{D\O\xspace}     


\newcommand{\de}{\ensuremath{^\circ}}
\newcommand{\ten}[1]{\ensuremath{\times \text{10}^\text{#1}}}
\newcommand{\unit}[1]{\ensuremath{\text{\,#1}}\xspace}
\newcommand{\mum}{\ensuremath{\,\mu\text{m}}\xspace}
\newcommand{\micron}{\ensuremath{\,\mu\text{m}}\xspace}
\newcommand{\cm}{\ensuremath{\,\text{cm}}\xspace}
\newcommand{\mm}{\ensuremath{\,\text{mm}}\xspace}
\newcommand{\mus}{\ensuremath{\,\mu\text{s}}\xspace}
\newcommand{\keV}{\ensuremath{\,\text{ke\hspace{-.08em}V}}\xspace}
\newcommand{\MeV}{\ensuremath{\,\text{Me\hspace{-.08em}V}}\xspace}
\newcommand{\GeV}{\ensuremath{\,\text{Ge\hspace{-.08em}V}}\xspace}
\newcommand{\TeV}{\ensuremath{\,\text{Te\hspace{-.08em}V}}\xspace}
\newcommand{\PeV}{\ensuremath{\,\text{Pe\hspace{-.08em}V}}\xspace}
\newcommand{\keVc}{\ensuremath{{\,\text{ke\hspace{-.08em}V\hspace{-0.16em}/\hspace{-0.08em}c}}}\xspace}
\newcommand{\MeVc}{\ensuremath{{\,\text{Me\hspace{-.08em}V\hspace{-0.16em}/\hspace{-0.08em}c}}}\xspace}
\newcommand{\GeVc}{\ensuremath{{\,\text{Ge\hspace{-.08em}V\hspace{-0.16em}/\hspace{-0.08em}c}}}\xspace}
\newcommand{\TeVc}{\ensuremath{{\,\text{Te\hspace{-.08em}V\hspace{-0.16em}/\hspace{-0.08em}c}}}\xspace}
\newcommand{\keVcc}{\ensuremath{{\,\text{ke\hspace{-.08em}V\hspace{-0.16em}/\hspace{-0.08em}c}^\text{2}}}\xspace}
\newcommand{\MeVcc}{\ensuremath{{\,\text{Me\hspace{-.08em}V\hspace{-0.16em}/\hspace{-0.08em}c}^\text{2}}}\xspace}
\newcommand{\GeVcc}{\ensuremath{{\,\text{Ge\hspace{-.08em}V\hspace{-0.16em}/\hspace{-0.08em}c}^\text{2}}}\xspace}
\newcommand{\TeVcc}{\ensuremath{{\,\text{Te\hspace{-.08em}V\hspace{-0.16em}/\hspace{-0.08em}c}^\text{2}}}\xspace}

\newcommand{\pbinv} {\mbox{\ensuremath{\,\text{pb}^\text{$-$1}}}\xspace}
\newcommand{\fbinv} {\mbox{\ensuremath{\,\text{fb}^\text{$-$1}}}\xspace}
\newcommand{\nbinv} {\mbox{\ensuremath{\,\text{nb}^\text{$-$1}}}\xspace}
\newcommand{\percms}{\ensuremath{\,\text{cm}^\text{$-$2}\,\text{s}^\text{$-$1}}\xspace}
\newcommand{\lumi}{\ensuremath{\mathcal{L}}\xspace}
\newcommand{\Lumi}{\ensuremath{\mathcal{L}}\xspace}
%
\newcommand{\LvLow}  {\ensuremath{\mathcal{L}=\text{10}^\text{32}\,\text{cm}^\text{$-$2}\,\text{s}^\text{$-$1}}\xspace}
\newcommand{\LLow}   {\ensuremath{\mathcal{L}=\text{10}^\text{33}\,\text{cm}^\text{$-$2}\,\text{s}^\text{$-$1}}\xspace}
\newcommand{\lowlumi}{\ensuremath{\mathcal{L}=\text{2}\times \text{10}^\text{33}\,\text{cm}^\text{$-$2}\,\text{s}^\text{$-$1}}\xspace}
\newcommand{\LMed}   {\ensuremath{\mathcal{L}=\text{2}\times \text{10}^\text{33}\,\text{cm}^\text{$-$2}\,\text{s}^\text{$-$1}}\xspace}
\newcommand{\LHigh}  {\ensuremath{\mathcal{L}=\text{10}^\text{34}\,\text{cm}^\text{$-$2}\,\text{s}^\text{$-$1}}\xspace}
\newcommand{\hilumi} {\ensuremath{\mathcal{L}=\text{10}^\text{34}\,\text{cm}^\text{$-$2}\,\text{s}^\text{$-$1}}\xspace}


\newcommand{\zp}{\ensuremath{\mathrm{Z}^\prime}\xspace}


\newcommand{\kt}{\ensuremath{k_{\mathrm{T}}}\xspace}
\newcommand{\BC}{\ensuremath{{B_{\mathrm{c}}}}\xspace}
\newcommand{\bbarc}{\ensuremath{{\overline{b}c}}\xspace}
\newcommand{\bbbar}{\ensuremath{{b\overline{b}}}\xspace}
\newcommand{\ccbar}{\ensuremath{{c\overline{c}}}\xspace}
\newcommand{\JPsi}{\ensuremath{{J}/\psi}\xspace}
\newcommand{\bspsiphi}{\ensuremath{B_s \to \JPsi\, \phi}\xspace}
\newcommand{\AFB}{\ensuremath{A_\mathrm{FB}}\xspace}
\newcommand{\EE}{\ensuremath{e^+e^-}\xspace}
\newcommand{\MM}{\ensuremath{\mu^+\mu^-}\xspace}
\newcommand{\TT}{\ensuremath{\tau^+\tau^-}\xspace}
\newcommand{\wangle}{\ensuremath{\sin^{2}\theta_{\mathrm{eff}}^\mathrm{lept}(M^2_\mathrm{Z})}\xspace}
\newcommand{\ttbar}{\ensuremath{{t\overline{t}}}\xspace}
\newcommand{\stat}{\ensuremath{\,\text{(stat.)}}\xspace}
\newcommand{\syst}{\ensuremath{\,\text{(syst.)}}\xspace}

\newcommand{\HGG}{\ensuremath{\mathrm{H}\to\gamma\gamma}}
\newcommand{\gev}{\GeV}
\newcommand{\GAMJET}{\ensuremath{\gamma + \mathrm{jet}}}
\newcommand{\PPTOJETS}{\ensuremath{\mathrm{pp}\to\mathrm{jets}}}
\newcommand{\PPTOGG}{\ensuremath{\mathrm{pp}\to\gamma\gamma}}
\newcommand{\PPTOGAMJET}{\ensuremath{\mathrm{pp}\to\gamma +
\mathrm{jet}
}}
\newcommand{\MH}{\ensuremath{\mathrm{M_{\mathrm{H}}}}}
\newcommand{\RNINE}{\ensuremath{\mathrm{R}_\mathrm{9}}}
\newcommand{\DR}{\ensuremath{\Delta\mathrm{R}}}


\newcommand{\PT}{\ensuremath{p_{\mathrm{T}}}\xspace}
\newcommand{\pt}{\ensuremath{p_{\mathrm{T}}}\xspace}
\newcommand{\ET}{\ensuremath{E_{\mathrm{T}}}\xspace}
\newcommand{\HT}{\ensuremath{H_{\mathrm{T}}}\xspace}
\newcommand{\et}{\ensuremath{E_{\mathrm{T}}}\xspace}
\newcommand{\Em}{\ensuremath{E\!\!\!/}\xspace}
\newcommand{\Pm}{\ensuremath{p\!\!\!/}\xspace}
\newcommand{\PTm}{\ensuremath{{p\!\!\!/}_{\mathrm{T}}}\xspace}
\newcommand{\ETm}{\ensuremath{E_{\mathrm{T}}^{\mathrm{miss}}}\xspace}
\newcommand{\MET}{\ensuremath{E_{\mathrm{T}}^{\mathrm{miss}}}\xspace}
\newcommand{\ETmiss}{\ensuremath{E_{\mathrm{T}}^{\mathrm{miss}}}\xspace}
\newcommand{\VEtmiss}{\ensuremath{{\vec E}_{\mathrm{T}}^{\mathrm{miss}}}\xspace}

%

\newcommand{\ga}{\ensuremath{\gtrsim}}
\newcommand{\la}{\ensuremath{\lesssim}}
\newcommand{\swsq}{\ensuremath{\sin^2\theta_W}\xspace}
\newcommand{\cwsq}{\ensuremath{\cos^2\theta_W}\xspace}
\newcommand{\tanb}{\ensuremath{\tan\beta}\xspace}
\newcommand{\tanbsq}{\ensuremath{\tan^{2}\beta}\xspace}
\newcommand{\sidb}{\ensuremath{\sin 2\beta}\xspace}
\newcommand{\alpS}{\ensuremath{\alpha_S}\xspace}
\newcommand{\alpt}{\ensuremath{\tilde{\alpha}}\xspace}

\newcommand{\QL}{\ensuremath{Q_L}\xspace}
\newcommand{\sQ}{\ensuremath{\tilde{Q}}\xspace}
\newcommand{\sQL}{\ensuremath{\tilde{Q}_L}\xspace}
\newcommand{\ULC}{\ensuremath{U_L^C}\xspace}
\newcommand{\sUC}{\ensuremath{\tilde{U}^C}\xspace}
\newcommand{\sULC}{\ensuremath{\tilde{U}_L^C}\xspace}
\newcommand{\DLC}{\ensuremath{D_L^C}\xspace}
\newcommand{\sDC}{\ensuremath{\tilde{D}^C}\xspace}
\newcommand{\sDLC}{\ensuremath{\tilde{D}_L^C}\xspace}
\newcommand{\LL}{\ensuremath{L_L}\xspace}
\newcommand{\sL}{\ensuremath{\tilde{L}}\xspace}
\newcommand{\sLL}{\ensuremath{\tilde{L}_L}\xspace}
\newcommand{\ELC}{\ensuremath{E_L^C}\xspace}
\newcommand{\sEC}{\ensuremath{\tilde{E}^C}\xspace}
\newcommand{\sELC}{\ensuremath{\tilde{E}_L^C}\xspace}
\newcommand{\sEL}{\ensuremath{\tilde{E}_L}\xspace}
\newcommand{\sER}{\ensuremath{\tilde{E}_R}\xspace}
\newcommand{\sFer}{\ensuremath{\tilde{f}}\xspace}
\newcommand{\sQua}{\ensuremath{\tilde{q}}\xspace}
\newcommand{\sUp}{\ensuremath{\tilde{u}}\xspace}
\newcommand{\suL}{\ensuremath{\tilde{u}_L}\xspace}
\newcommand{\suR}{\ensuremath{\tilde{u}_R}\xspace}
\newcommand{\sDw}{\ensuremath{\tilde{d}}\xspace}
\newcommand{\sdL}{\ensuremath{\tilde{d}_L}\xspace}
\newcommand{\sdR}{\ensuremath{\tilde{d}_R}\xspace}
\newcommand{\sTop}{\ensuremath{\tilde{t}}\xspace}
\newcommand{\stL}{\ensuremath{\tilde{t}_L}\xspace}
\newcommand{\stR}{\ensuremath{\tilde{t}_R}\xspace}
\newcommand{\stone}{\ensuremath{\tilde{t}_1}\xspace}
\newcommand{\sttwo}{\ensuremath{\tilde{t}_2}\xspace}
\newcommand{\sBot}{\ensuremath{\tilde{b}}\xspace}
\newcommand{\sbL}{\ensuremath{\tilde{b}_L}\xspace}
\newcommand{\sbR}{\ensuremath{\tilde{b}_R}\xspace}
\newcommand{\sbone}{\ensuremath{\tilde{b}_1}\xspace}
\newcommand{\sbtwo}{\ensuremath{\tilde{b}_2}\xspace}
\newcommand{\sLep}{\ensuremath{\tilde{l}}\xspace}
\newcommand{\sLepC}{\ensuremath{\tilde{l}^C}\xspace}
\newcommand{\sEl}{\ensuremath{\tilde{e}}\xspace}
\newcommand{\sElC}{\ensuremath{\tilde{e}^C}\xspace}
\newcommand{\seL}{\ensuremath{\tilde{e}_L}\xspace}
\newcommand{\seR}{\ensuremath{\tilde{e}_R}\xspace}
\newcommand{\snL}{\ensuremath{\tilde{\nu}_L}\xspace}
\newcommand{\sMu}{\ensuremath{\tilde{\mu}}\xspace}
\newcommand{\sNu}{\ensuremath{\tilde{\nu}}\xspace}
\newcommand{\sTau}{\ensuremath{\tilde{\tau}}\xspace}
\newcommand{\Glu}{\ensuremath{g}\xspace}
\newcommand{\sGlu}{\ensuremath{\tilde{g}}\xspace}
\newcommand{\Wpm}{\ensuremath{W^{\pm}}\xspace}
\newcommand{\sWpm}{\ensuremath{\tilde{W}^{\pm}}\xspace}
\newcommand{\Wz}{\ensuremath{W^{0}}\xspace}
\newcommand{\sWz}{\ensuremath{\tilde{W}^{0}}\xspace}
\newcommand{\sWino}{\ensuremath{\tilde{W}}\xspace}
\newcommand{\Bz}{\ensuremath{B^{0}}\xspace}
\newcommand{\sBz}{\ensuremath{\tilde{B}^{0}}\xspace}
\newcommand{\sBino}{\ensuremath{\tilde{B}}\xspace}
\newcommand{\Zz}{\ensuremath{Z^{0}}\xspace}
\newcommand{\sZino}{\ensuremath{\tilde{Z}^{0}}\xspace}
\newcommand{\sGam}{\ensuremath{\tilde{\gamma}}\xspace}
\newcommand{\chiz}{\ensuremath{\tilde{\chi}^{0}}\xspace}
\newcommand{\chip}{\ensuremath{\tilde{\chi}^{+}}\xspace}
\newcommand{\chim}{\ensuremath{\tilde{\chi}^{-}}\xspace}
\newcommand{\chipm}{\ensuremath{\tilde{\chi}^{\pm}}\xspace}
\newcommand{\Hone}{\ensuremath{H_{d}}\xspace}
\newcommand{\sHone}{\ensuremath{\tilde{H}_{d}}\xspace}
\newcommand{\Htwo}{\ensuremath{H_{u}}\xspace}
\newcommand{\sHtwo}{\ensuremath{\tilde{H}_{u}}\xspace}
\newcommand{\sHig}{\ensuremath{\tilde{H}}\xspace}
\newcommand{\sHa}{\ensuremath{\tilde{H}_{a}}\xspace}
\newcommand{\sHb}{\ensuremath{\tilde{H}_{b}}\xspace}
\newcommand{\sHpm}{\ensuremath{\tilde{H}^{\pm}}\xspace}
\newcommand{\hz}{\ensuremath{h^{0}}\xspace}
\newcommand{\Hz}{\ensuremath{H^{0}}\xspace}
\newcommand{\Az}{\ensuremath{A^{0}}\xspace}
\newcommand{\Hpm}{\ensuremath{H^{\pm}}\xspace}
\newcommand{\sGra}{\ensuremath{\tilde{G}}\xspace}
\newcommand{\mtil}{\ensuremath{\tilde{m}}\xspace}
\newcommand{\rpv}{\ensuremath{\rlap{\kern.2em/}R}\xspace}
\newcommand{\LLE}{\ensuremath{LL\bar{E}}\xspace}
\newcommand{\LQD}{\ensuremath{LQ\bar{D}}\xspace}
\newcommand{\UDD}{\ensuremath{\overline{UDD}}\xspace}
\newcommand{\Lam}{\ensuremath{\lambda}\xspace}
\newcommand{\Lamp}{\ensuremath{\lambda'}\xspace}
\newcommand{\Lampp}{\ensuremath{\lambda''}\xspace}
\newcommand{\spinbd}[2]{\ensuremath{\bar{#1}_{\dot{#2}}}\xspace}

\newcommand{\MD}{\ensuremath{{M_\mathrm{D}}}\xspace}
\newcommand{\Mpl}{\ensuremath{{M_\mathrm{Pl}}}\xspace}
\newcommand{\Rinv} {\ensuremath{{R}^{-1}}\xspace}

%
%
\hyphenation{en-viron-men-tal}

%% file: CFT-09-008-authorlist.tex
\textbf{Yerevan Physics Institute,  Yerevan,  Armenia}\\*[0pt]
S.~Chatrchyan, V.~Khachatryan, A.M.~Sirunyan
\vskip\cmsinstskip
\textbf{Institut f\"{u}r Hochenergiephysik der OeAW,  Wien,  Austria}\\*[0pt]
W.~Adam, B.~Arnold, H.~Bergauer, T.~Bergauer, M.~Dragicevic, M.~Eichberger, J.~Er\"{o}
% ...
\vskip\cmsinstskip
% ...
